\documentclass[graybox,english,11pt]{article}
\usepackage{geometry}
\usepackage[latin9]{inputenc}
\usepackage{array}
\usepackage{verbatim}
\usepackage{float}
\usepackage{bm}
\usepackage{multirow}
\usepackage{amsthm}
\usepackage{amsmath}
\usepackage{amssymb}
\usepackage{graphicx}
\usepackage{color}
\usepackage{appendix}
\usepackage{subfig}
\usepackage{epstopdf}

\makeatletter

\theoremstyle{plain}

 \theoremstyle{definition}
  
  \theoremstyle{plain}
  
  \theoremstyle{plain}

  \theoremstyle{remark}
  \newtheorem*{rem*}{\protect\remarkname}
  \theoremstyle{plain}
  
  \theoremstyle{remark}
  
  \theoremstyle{plain}
  
 \theoremstyle{plain}
  \newtheorem*{theorem*}{\protect\theoremname}
\theoremstyle{plain}
  
\makeatother

\usepackage[english]{babel}
  \providecommand{\algorithmname}{Algorithm}
  \providecommand{\assumptionname}{Assumption}
  \providecommand{\lemmaname}{Lemma}
  \providecommand{\problemname}{Problem}
  \providecommand{\propositionname}{Proposition}
  \providecommand{\remarkname}{Remark}
\providecommand{\theoremname}{Theorem}
\providecommand{\definitionname}{Definition}
\providecommand{\notationname}{Notation}

\usepackage[bottom]{footmisc}
\usepackage{graphics}

\newcommand{\R}{\mathbb{R}}

\usepackage{color}

\begin{document}

\title{Conservative, Dissipative and Super-diffusive Behavior of a Particle Propelled in a Regular Flow}
\author{Gil Ariel\thanks{Department of Mathematics, Bar-Ilan University,
Ramat-Gan 5290002, Israel. (arielg@math.biu.ac.il)}
\and Jeremy Schiff\thanks{Department of Mathematics, Bar-Ilan University,
Ramat-Gan 5290002, Israel. (schiff@math.biu.ac.il)}}

\date{\today}

\maketitle

\begin{abstract}
A recent model of Ariel {\em et al.} \cite{ArielPRL2017} for explaining the
observation of L\'evy walks in swarming bacteria suggests that
self-propelled, elongated particles in a periodic array of regular vortices 
perform a super-diffusion that is consistent with L\'evy walks.
The equations of motion, which are reversible in time but not volume preserving, 
demonstrate a new route to L\'evy walking in chaotic systems.
Here, the dynamics of the model is studied both analytically and numerically. 
It is shown that the apparent super-diffusion is due to ``sticking'' of
trajectories to  elliptic islands, regions of quasi-periodic orbits 
reminiscent of those seen in conservative systems. However, for certain  
parameter values, these islands coexist with asymptotically stable periodic 
trajectories, causing dissipative behavior on very long time scales. 
\end{abstract}


\section{Introduction} \label{sec:introduction}

L\'evy walks are random process that are characterized by trajectories that 
have straight stretches for extended lengths whose variance is infinite. 
It is well known that L\'evy walks can arise in chaotic systems \cite{Klafter1996}.
Indeed, the literature contains a potpourri of models and theories explaining 
the mechanisms underlying such dynamics.
Notable examples include Hamiltonian systems 
\cite{Geisel1985,Geisel1987,Zaslavsky1993,Zaslavsky1995}, 
area-preserving maps \cite{Meiss1985,Meiss1986,Yu1990},
chaotic or random velocity fields \cite{Bouchaud1990}, 
time varying fields in resonance \cite{Aranson1990,Mesquita1992,Solomon2001}, 
dissipative systems \cite{PomeauManneville1980},
turbulence \cite{Brandstater1987} and more \cite{Zaburdeav2015}.

\begin{figure}[t]
   \centerline{
   \includegraphics[width=0.5\textwidth]{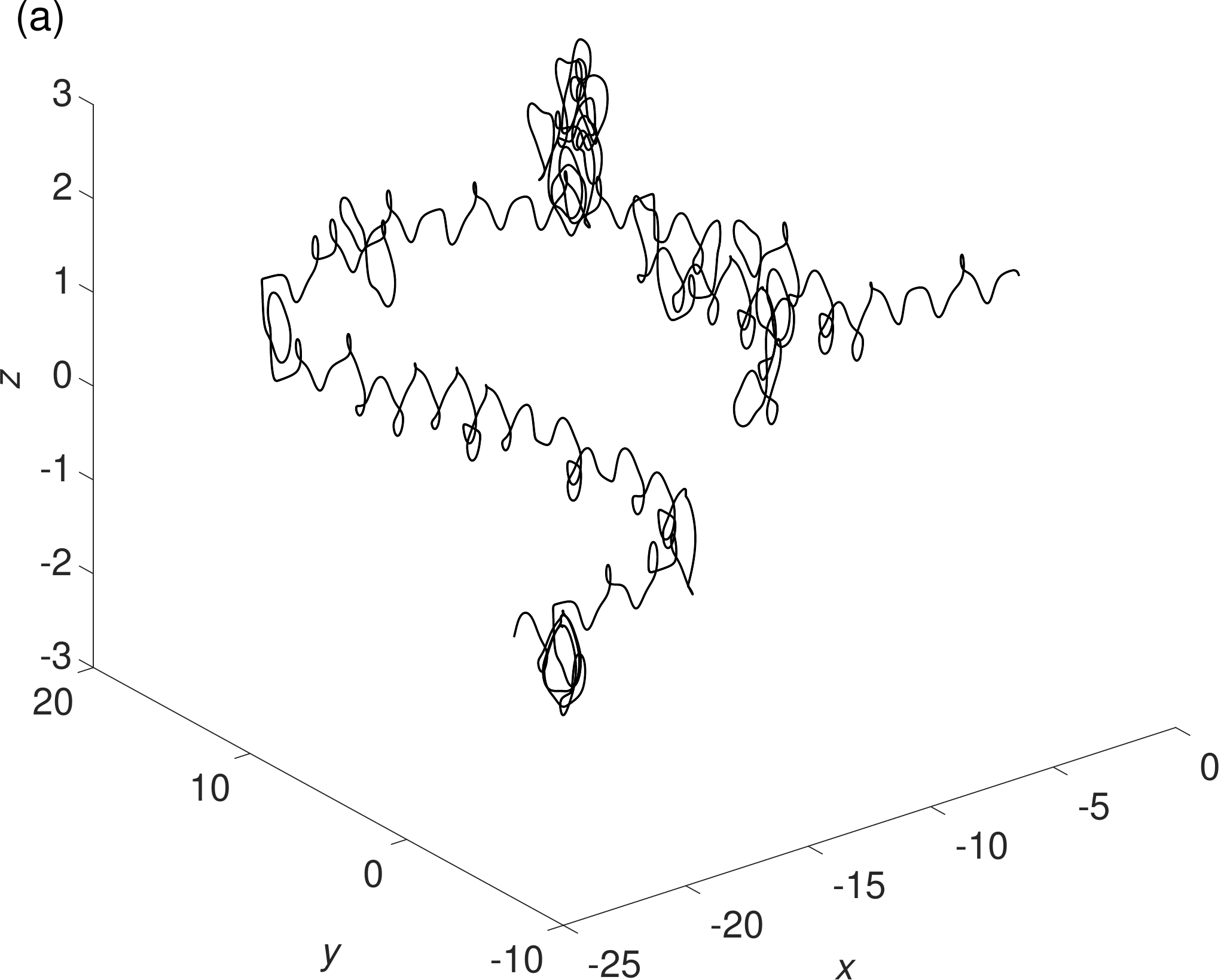}~~~~~~
   \includegraphics[width=0.4\textwidth]{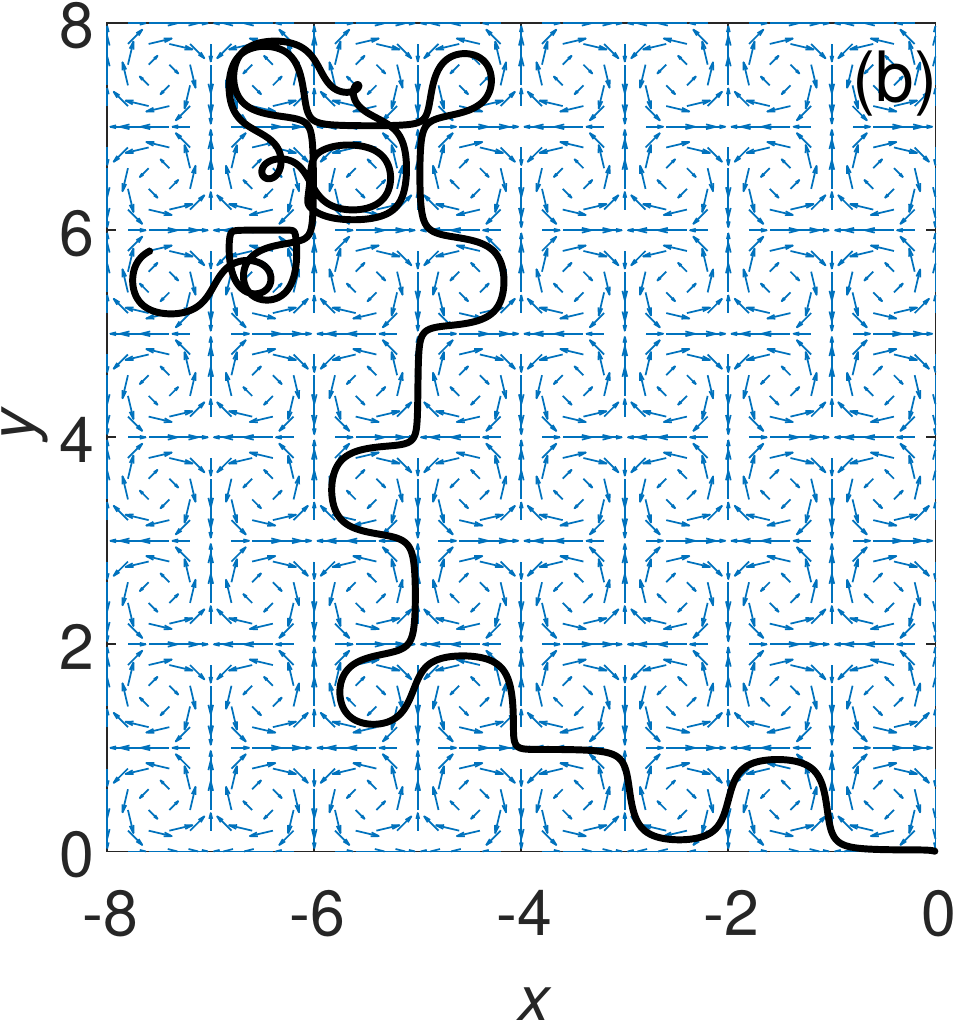}
   }
\caption{(a) An example trajectory, with  
$V=1/2$, $D=12/13$ and initial conditions
$(x(0),y(0),z(0))=(0,0,0.9)$.
(b) Projection  of part of the trajectory
onto the $x$-$y$ plane, with the vortex field  displayed.}
\label{fig:exampleTrajectory} 
\end{figure}

Here, we consider the following set of ordinary differential equations (ODEs), describing the position and orientation of 
an infinitesimal, self-propelled, prolate spheroid that is advected by an 
array of vortices: 
\begin{equation}
\begin{aligned}
   &\dot{x}  = \sin \pi x \cos \pi y + V   \cos \pi z\ , \\ 
   &\dot{y} =  -\cos \pi x \sin \pi y  + V  \sin \pi z\ ,  \\
   &\dot{z} = \sin \pi x \sin \pi y 
          - 2D \cos \pi x \cos \pi y \cos \pi z \sin \pi z\ . 
\end{aligned}
\label{eq:model}
\end{equation}
Here $x$ and $y$ describe the position of a point particle moving in the 
plane and advected along an array of vortices 
with flow lines $\psi(x,y)=\pi^{-1} \sin \pi x \sin \pi y$. 
Assuming Stokes dynamics, the particle pushes itself at constant speed 
$V \in [0,1]$ in the direction of its ``head'' with angle $\pi z$ 
relative to the $x$-axis. 
The rotation of the head is given by Jeffery's equation, 
where $D \in [0,1]$ is a shape parameter.
Figure \ref{fig:exampleTrajectory} shows (a) a typical trajectory and 
(b) projection of part of the trajectory 
onto the $x$-$y$ plane, on which the vortex field is displayed. 
The ODEs \eqref{eq:model} have been considered by Torney and Neufeld
\cite{TorneyNeufeld2013} and independently by Ariel {\it et al.} 
\cite{ArielPRL2017} as a simplified model for explaining the observation 
of L\'evy walks in swarming bacteria \cite{ArielNatureComm2015}.
In the following we focus on the dynamics of \eqref{eq:model} in order 
to better understand the mechanisms resulting in super-diffusive trajectories.

This paper is organized as follows. In  section \ref{sec:analysis} 
we go through the basic analytic properties of the system 
\eqref{eq:model} and in particular its symmetries. The system has a 
time reversal symmetry, which has important implications for the dynamics. 
We discuss periodic  orbits, and give a numeric-aided proof 
of the existence of a periodic orbit in  a specific case. We show this orbit is 
surrounded by an elliptic island, similar to those observed in conservative 
systems. The  ``stickiness'' of  such  islands (i.e. the fact that nearby orbits 
can get stuck close to the island for long periods of time) seems to be 
a significant mechanism underlying super-diffusivity.  In 
section \ref{sec:numerics} numeric results are presented. A specialized 
numerical method is described that gives accurate 
solutions over extended time scales. This is used to investigate 
the dependence on the parameters $D$ and $V$ of various properties of the system, 
such as the exponent of super-diffusivity.  Rather surprisingly, a
region of parameter space is identified in which the numerics indicate that 
this exponent has the constant value $2$, corresponding to ballistic motion
(also called accelerator modes \cite{HC}). Further exploration shows that in this
region there exist attractive periodic orbits, and suggests that for some 
parameter values, almost 
all orbits outside the elliptic islands tend to these orbits, but only after
very long times. For other parameter values, however, only orbits sufficiently close
to the attractive orbits are attracted to them. 
We conclude in Section~\ref{sec:summary}.

\section{Basic analysis}   \label{sec:analysis}
\setcounter{equation}{0}

First we note that the system is not Hamiltonian, as the number of variables is odd. Indeed the dynamics is not even volume preserving, as the divergence of the right hand side of    \eqref{eq:model} is 
$ - 2 \pi D \cos \pi x \cos \pi y \cos 2 \pi z$, which is typically not zero.

Next, it is easily verified that the ODE system \eqref{eq:model} is invariant under the symmetry $(x,y,z) \mapsto (x+2,y+2,z+2)$. Thus the dynamics can be 
thought of as occurring in the $3$-dimensional torus $\Omega=[-1,1]^3$.
Additional symmetries are  
\begin{itemize}
\item $(x,y,z) \mapsto (-x,y,1-z)$
\item $(x,y,z) \mapsto (x,-y,-z)$
\item $(x,y,z) \mapsto (x+1,y+1,z)$
\end{itemize}
and compositions of the above.
Hence the basic domain $\Omega$ is further partitioned into 8 symmetric parts. 

Most importantly, the dynamics is {\em reversible}, in the sense that there exists
an involution $G$ of $\Omega$ (i.e. a map from $\Omega$ to itself with 
$G^2=I$) such that solving the ODE backward in time is the same as applying $G$, 
solving forward in time, and then applying $G$ again. 
If we denote the propagator of \eqref{eq:model} as $\Phi_t $ (i.e 
$\Phi_t(x_0,y_0,z_0)$ is the solution of \eqref{eq:model} with 
initial condition $(x_0,y_0,z_0)$), 
then  the dynamics is reversible if $\Phi_t G \Phi_t G = 1$.
It is easily verified that taking
\begin{equation}
	G : (x,y,z) \mapsto \left(y,x,\frac32 - z\right)
\label{eq:G}
\end{equation}
satisfies this requirement.
Reversible dynamical systems can exhibit types of behavior associated
with conservative systems, specifically they can have ``elliptic islands'' 
in which a periodic elliptic orbit is surrounded by invariant surfaces,
between which there may be other periodic orbits, surrounded by other invariant 
surfaces, and so on. However, they can also exhibit features of dissipative 
systems, such as attractors and repellors. See, for example, the review by Roberts and Quispel \cite{RobertsQuispel1992}, for the case of reversible 
maps of the plane, and the two more recent works \cite{delshams,turaev}. 
In a reversible system, fixed points and periodic solutions can be either 
{\em symmetric} (i.e. mapped to themselves by the involution $G$) or 
{\em asymmetric}.  Clearly, attractors  and repellors cannot be 
symmetric, as $G$ must map an attractor to a repellor and vice-versa. 

The first step in the investigation of the dynamics of any system is 
to find the fixed points, and try to spot simple invariant manifolds. For 
 \eqref{eq:model}, it is straightforward to check that there are $16$
fixed points, two of these being the points with $y=z=0$, $\sin \pi x = -V$,
and the remaining $14$ being generated from these by application of 
the three symmetries and the time reversing involution $G$ described above.
Looking at the linearizations  near 
the $16$ fixed points, $8$ have two eigenvalues with positive real part
and one with negative real part, and the remaining $8$ have two eigenvalues with negative real part and one with positive real part. The eigenvalues are 
all real if $\left(1+\left(D-\frac12\right)^2\right)(1-V^2)\ge 1$, otherwise
there is a complex conjugate pair.  With regard to invariant manifolds, the 
circle $y=z=0$ is invariant, as are  $7$ other
circles obtained by the applying the symmetries and $G$. Each of these 
circles passes through two of the fixed points (one with a two-dimensional 
stable manifold, one with a two-dimensional unstable manifold), and, 
apart from these points, consists of a pair of heteroclinic orbits. 

%

\begin{figure}[t]
   \centerline{   \includegraphics[width=0.4\textwidth]{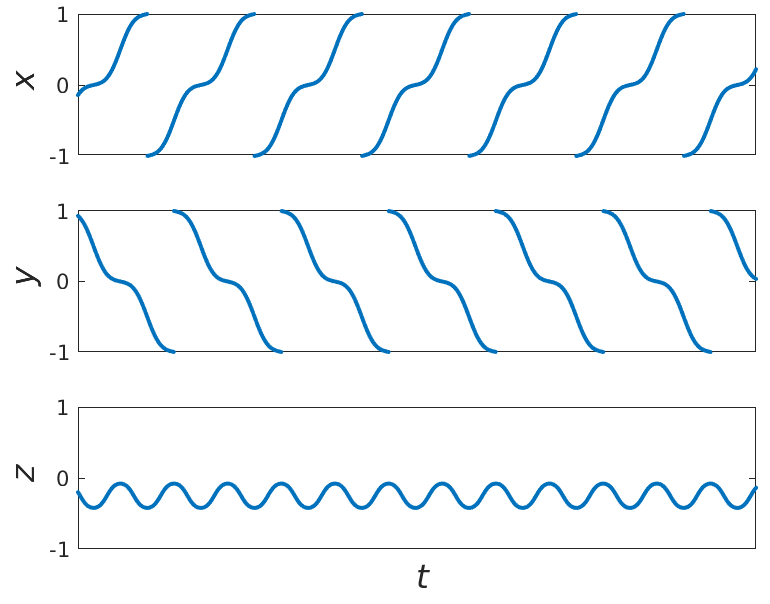} ~~~~~
                   \includegraphics[width=0.55\textwidth]{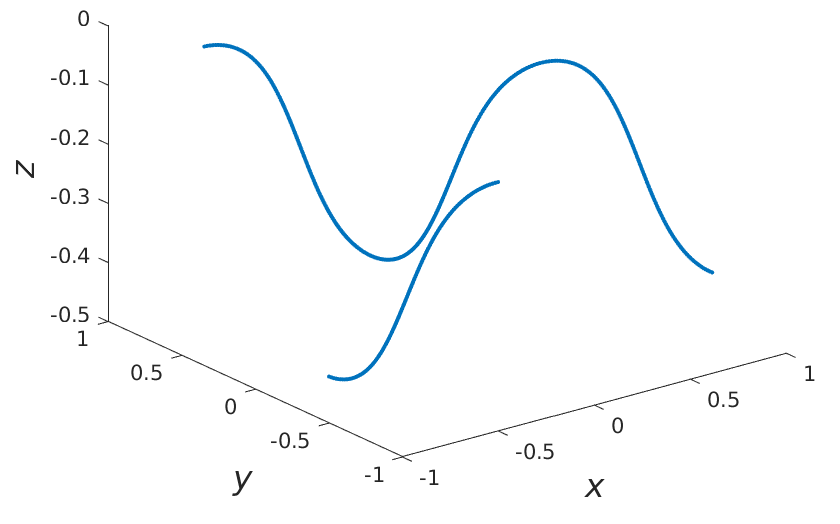}}
   \caption{ A periodic solution in the case $D=12/13$, $V=1/2$, 
   passing  approximately through the point $(1.858624,0.930363,-0.2)$.
   Left: $x,y,z$ as functions of $t$. Right: 3-d view of the orbit.} 
\label{fig:T1} 
\end{figure}

The next question to consider is whether there exist periodic orbits. There is, in
general, no analytic way to resolve this, but numerical experiments show that 
the system  \eqref{eq:model}  does indeed have many periodic orbits. Note that 
for  a general  flow $\dot{\bf x}=F({\bf x})$  on the torus 
$[-1,1]^d$ a periodic orbit takes the form 
\begin{equation}
{\bf x}(t) = \frac{2{\bf n}t}{T} + {\bf y}(t) ,
\end{equation}
where ${\bf y}:{\bf R}\rightarrow{\bf R}^d$ is periodic with 
period $T$ and ${\bf n}\in{\bf Z}^d$. 
Thus, for the system \eqref{eq:model}, each periodic orbit is characterized 
by $3$ integers, which specify the 
number of times the orbit winds around the torus in each of the $x,y,z$ 
directions in  the course of a single period.  In the following we take 
parameter values $D=12/13$ and $V=1/2$, which 
are values that are realistic for the spreading phenomenon of bacteria within a 
swarm \cite{ArielPRL2017}. For these parameters one periodic orbit
goes approximately through the point $(x_p,y_p,z_p)=(1.858624,0.930363,-0.2)$. This has 
period approximately $2.77$ and the winding numbers are $n_x=1,n_y=-1,n_z=0$. 
See figure \ref{fig:T1}. This orbit is invariant under $G$ (Note that $G$ 
changes the sign of both $n_x$ and $n_y$ while leaving $n_z$ invariant, thus 
a necessary condition for $G$ invariance is $n_x=-n_y$). 
The orbit is also invariant under one of the $3$ symmetries of the system, 
but applying the other two symmetries generates $3$ more periodic orbits, 
with winding numbers (1) $n_x=n_y=1,n_z=0$, (2) $n_x=n_y=-1, n_z=0$ and 
(3) $n_x=-1,n_y=1,n_z=0$.  

\begin{figure}[t]
\centerline{
   \includegraphics[width=0.6\textwidth]{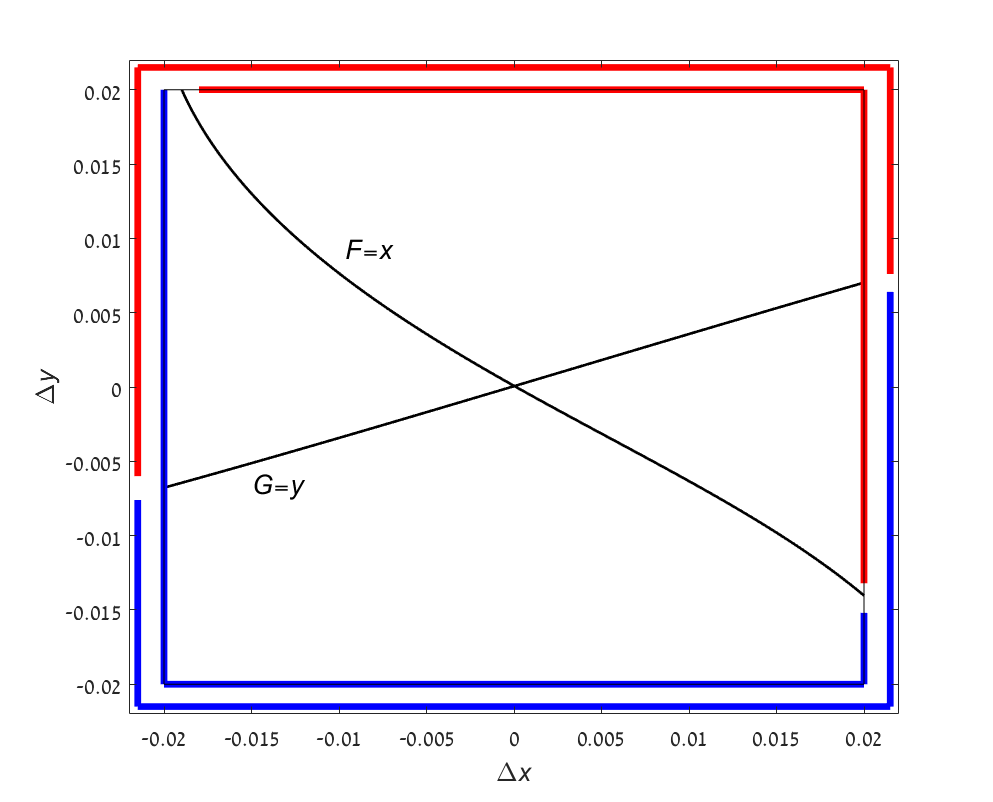}}
   \caption{ Proof of existence of periodic solutions.
   Outer square blue: $F(x,y)-x$ is positive. 
    Outer square red: $F(x,y)-x$ is negative. 
    Inner square blue: $G(x,y)-y$ is positive. 
    Inner square red: $G(x,y)-y$ is negative. The zero level set of $F-x$ connects
    some point in the gap on the top of the outer square with some point in 
    the gap on the right side of the outer square. The zero level set of $G-y$ connects
    some point in the gap on the left side of the inner square with some point in 
    the gap on the right side of the inner square. Clearly they must intersect 
    in the interior.}
\label{fig:TN} 
\end{figure}

The periodic orbit presented in the previous paragraph was found by looking 
at the return map to a surface of section, and we also use this to give 
a numeric-aided proof of existence of the orbit and, in the next paragraph, 
to study its stability. 
Specifically, we use the surface of section $z=-0.2$. We see that the orbit hits 
this 
$4$ times in a period, once in each quadrant of the $x$,$y$-plane, so we restrict
to the quadrant $-1<x<0$, $0<y<1$. Denote by $(F(x,y),G(x,y))$ the first return
map to this section, starting from the point $(x,y)$. Thus we seek a solution 
of 
\begin{equation}
\begin{aligned}
   x  &= F(x,y) \\ 
   y  &= G(x,y) .
\end{aligned}
\label{eq:returnMap}
\end{equation}
Consider a square domain $A$ on the surface of section in which the first 
return time is bounded by a constant $\tau$.
Denote the RHS of \eqref{eq:model} as $H(x,y,z)$.
Then, the derivatives of $F$ and $G$ are bounded by $\tau \max_A | \nabla H|$.
The accuracy of a numerical computation of $F$ and $G$ with a given integration 
scheme also depends on similar bounds.
The $n$'th order partial derivatives of $H$ are trivially bounded in the 
sup norm by $2(2\pi)^n$.
Hence, we can obtain strict bounds for both the accuracy of our numerical
computation of $F$ and $G$ and their derivatives.
Let ${\rm tol}>0$ denote a given accuracy tolerance. 
We compute $F$ and $G$ on a uniform grid along the boundary of $A$ and note
all points in which we can guarantee that 
$F(x,y)-x$ and $G(x,y)-y$ are strictly positive or negative 
(i.e. larger than tol or smaller than $-$tol).
The grid spacing should be small enough such that $F$ and $G$ cannot change by 
more than tol between neighboring points. 
Figure~\ref{fig:TN}  depicts our results with 
$A=[x_p-0.02,x_p+0.02] \times  [y_p-0.02,y_p+0.02]$
and tol=$10^{-3}$. The plot is shifted so that the fixed point is at the origin.
Blue (red) lines show positive (negative) values of $F-x$ (outer square) and 
$G-y$ 
(inner square). Segments along the boundary of $A$ which are not colored cannot 
be classified as positive or negative within the given precision.
As the figure shows, the region in which $F-x$ changes sign does not overlap that 
of $G-y$. Furthermore, following the boundary of $A$ in a fixed orientation (e.g., clock-wise), 
the regions in which $F-x$ and $G-y$ change sign alternate. It follows that 
the zero level curve of $F-x$ must intersect that of $G-y$ inside the square.
This intersection point is a fixed point for the return map and hence  
a  periodic solution of \eqref{eq:model}.

\begin{figure}[h!]
\centerline{\includegraphics[width=0.8\textwidth]{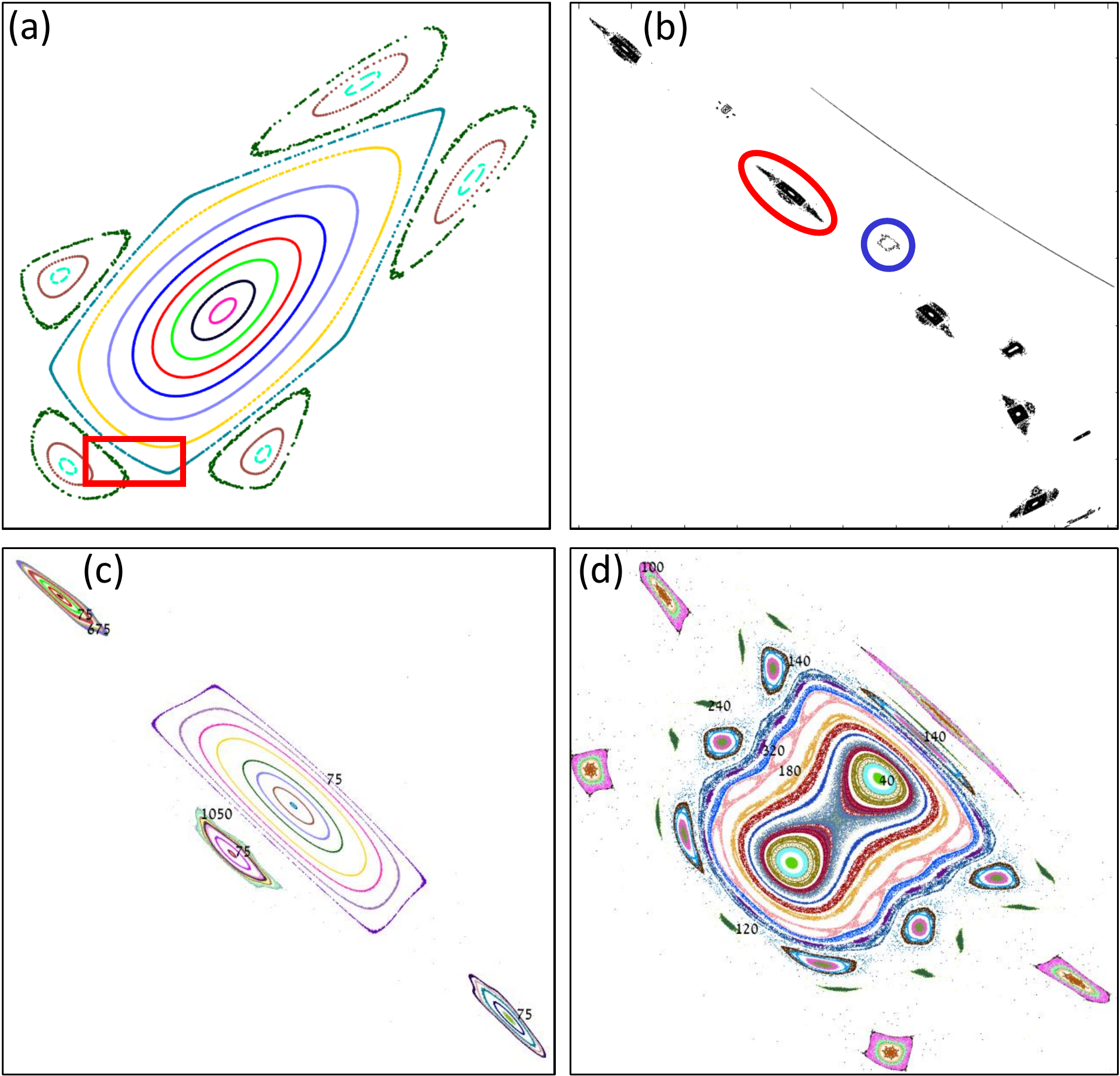}}
\caption{
Periodic orbits and their stability. All figures show the return map to the 
surface of section $z=-0.2$. 
Each color corresponds to a different initial condition.  
(a) The region $[1.75,2]\times [0.85,1.05]$ showing the T1 and  T5 
orbits. 
(b) A zoom into the red rectangle in (a) reveals additional features.
(c) A zoom into the region $[1.8108,1.8132]\times [0.8831,0.8848]$, including 
the red ellipse in (b). Numbers denote the period multiplier. 
(d) A zoom into the region $[1.8092,1.80995]\times [0.8862,0.88673]$, including
the blue circle in (b). }
\label{fig:fractals} 
\end{figure}

The stability of periodic orbits can be studied by numerically plotting the 
return map to the surface of section $z=-0.2$ near one of the periodic orbits, see 
Figure~\ref{fig:fractals}(a). Each color corresponds to a different initial condition.
The concentric ellipsoid-like curves are invariant curves
corresponding to quasiperiodic solutions, and their existence 
shows that the periodic solution corresponding to the fixed point in the 
center (which we call the T1 solution)
is neutral (i.e. stable but not asymptotically stable).
Further away from the center some of the invariant curves break up.  
Chaotic dynamics can be observed, interspersed with 
additional islands of stability around periodic orbits of longer period. 
Specifically in figure~\ref{fig:fractals}(a) we see five islands, 
surrounding an orbit we call T5, 
with period approximately 5 times that of T1.
Zooming in (see figures~\ref{fig:fractals}(b)-(d)), additional islands of 
stability can be seen, 
identified with  increasingly finer structures and longer periods. 
We identify periods up to 1050 times the period of T1.
At the center of the large island in figure~\ref{fig:fractals}(d) there is 
a hyperbolic fixed point. This phenomenon will be explained in  
section \ref{sec:GW}. 

\newpage

\begin{figure}[h!]
\centerline{\includegraphics[width=0.5\textwidth]{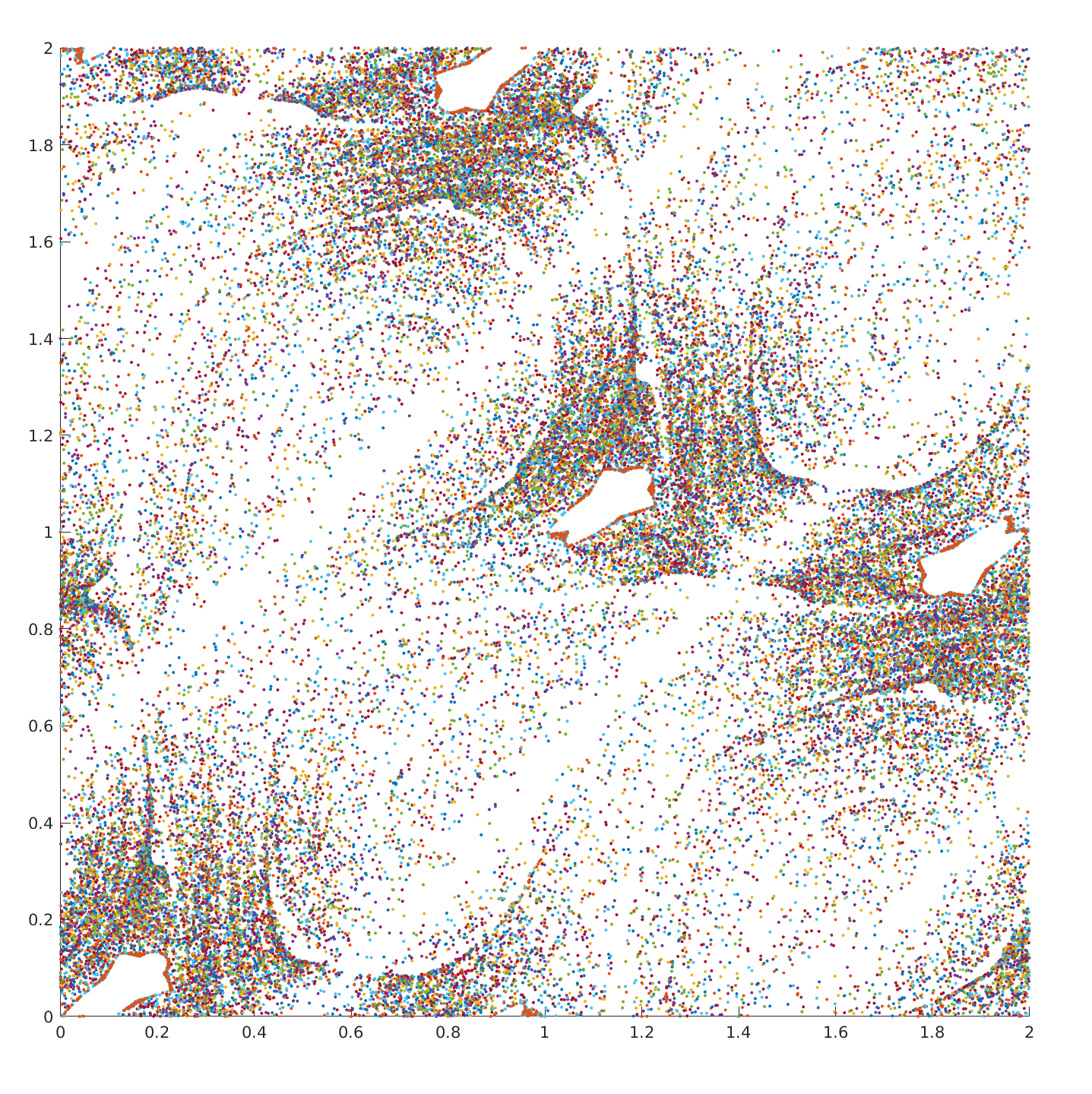}}
\caption{
Points on the return map to the 
surface of section $z=-0.2$ for $100$ orbits, run from $t=0$ until $t=2000$,
for $D=12/13$, $V=1/2$. 
The holes correspond to the regions of the islands of stability, which 
are ``sticky'' and have regions of high density close to them, corresponding to 
solutions that are close to ballistic for long periods of time. 
}
\label{fig:sticky} 
\end{figure}

Thus we see that the system has neutrally stable periodic orbits, and other 
solutions that remain close to, but are not attracted to, the periodic orbits.
How do generic orbits behave? In figure~\ref{fig:sticky} we show, for $D=12/13$,
$V=1/2$,  all points on the return map to the 
surface of section $z=-0.2$ for $100$ orbits, run from $t=0$ until $t=2000$.
The four ghost-like ``holes'' correspond to the regions of the islands of stability, 
one  of which is described in figure~\ref{fig:fractals}. None of the $100$ orbits 
were taken within these regions, which occupy approximately $1/50$ of the total area. 
However, we note that a large fraction of the points are found in certain regions
that,  roughly speaking, surround the holes. 
It is well known that trajectories can ``stick'' to the area close to regular regions 
\cite{RobertsQuispel1992}. When the sticking time has a power-law distribution, then 
super-diffusion and L\'evy walks can  occur. Loosely speaking, the reason is that 
while a  trajectory is close to a periodic or quasi-periodic
orbit, it  appears ballistic (in the unwrapped space).
The result is a L\'evy walk \cite{Klafter1996}.
The route from chaos to L\'evy walks has been mainly studied in the context of 
Hamiltonian or  area-preserving systems \cite{Meiss1985,Meiss1986,AlusFishman2014}.
Sticking has been attributed to long escape times from regions
that are bounded by broken KAM trajectories \cite{AlusFishman2014,AlusFishman2017}.
To the best of our knowledge, the origin of super-diffusion in the more general 
class of  reversible systems is not well understood \cite{RobertsQuispel1992}.

We note that in addition to the general accumulation of density in some 
regions close to the ``holes'', 
the plot in figure~\ref{fig:sticky} displays much additional structure, 
both in and away from the regions of high density. In the sequel, we refer
to the region  of phase space outside the islands of stability as the chaotic region 

\newpage 

\section{Long-time Numerics}
\label{sec:numerics}
\setcounter{equation}{0}

\subsection{A bespoke numerical method}

Estimation of the statistical properties of trajectories, for example 
the mean square displacement,  with reasonable accuracy, requires numerical
integration of  \eqref{eq:model} for long times.
Standard numerical methods, e.g. Runge-Kutta, are notoriously inappropriate 
for integrating chaotic systems, and  adaptive methods can become 
prohibitively inefficient. The problem is worse for long-time integration 
of equations such as \eqref{eq:model}.
Small noise due to numerical errors can make a trajectory jump from outside to inside one of the quasi-periodic regions described above, or vice-versa.
This leads to a qualitatively different geometry of trajectories. 

The antidote to these problems is to use an integrator that preserves 
the geometric properties of the system. Integrators that preserve symmetries,
including time reversal symmetries, have been developed, for example in 
\cite{mqt}. We take another approach. Elliptic islands 
are more commonly seen in Hamiltonian and volume-preserving systems. 
We show how the system  \eqref{eq:model} can be embedded 
(as motion on a 3-d invariant submanifold) in a 4-d 
volume preserving system, and apply a volume-preserving integrator to this. 
Although we find that it is necessary to project back the relevant 
3-d submanifold, we hope this method retains sufficient geometric structure to 
accurately integrate the system. 

Consider the 4-d system 
\begin{equation}
\begin{aligned}
   \dot{w} &= -\sin \pi x \sin \pi y + D \cos \pi x \cos \pi y \sin 2 \pi z \ ,\\
   \dot{x} &= \sin \pi x \cos \pi y + V \cos \pi w\ ,  \\
   \dot{y} &= - \cos \pi x \sin \pi y + V \sin \pi z\ ,  \\
   \dot{z} &= \sin \pi x \sin \pi y + D \cos \pi x \cos \pi y \sin 2 \pi w .
\end{aligned}
\label{eq:4D}
\end{equation}
This is obtained from \eqref{eq:model} by appending the equation for $w$ and 
replacing $z$ by $-w$ in two places. The RHS of this new system is now 
divergence free, and thus it is a volume preserving system. It is also 
straightforward to check that the surface $w=-z$ is invariant, and reduction 
to this surface gives the original system.  

Following an idea of Feng Kang \cite{FengKang1995}, we find three Hamiltonians such that the dynamics is a composition of three Hamiltonian maps.
There is some freedom in choosing how to split the system. 
Here, since the convection of particles along the flow lines is divergence free, we keep $x$ and $y$ as a pair. 
Similarly, for symmetry reasons, we pair $x$ with $w$ and $y$ with $z$.
Overall, we seek three Hamiltonians,
 $H_{xy}$, $H_{wx}$ and $H_{zy}$ such that
\begin{equation}
\begin{aligned}
   \dot{x} = \frac{\partial H_{xy}}{\partial y} + \frac{\partial H_{wx}}{\partial w},~~~ &   
   \dot{y} = -\frac{\partial H_{xy}}{\partial x} + \frac{\partial H_{zy}}{\partial z} \\
   \dot{w} =  -\frac{\partial H_{wx}}{\partial x},~~~ &
   \dot{z} =  -\frac{\partial H_{zy}}{\partial y}
\end{aligned}
\label{eq:HamiltonianDynamics}
\end{equation}
We find that a convenient choice is given by,
\begin{equation}
\begin{aligned}
   H_{xy} &=\pi^{-1} \sin \pi x \sin \pi y\ , \\
   H_{wx} &= \pi^{-1} [ V \sin \pi w - \cos \pi x \sin \pi y - D \sin \pi x \cos \pi y \sin 2 \pi z  ]\ ,\\
   H_{zy} &= -\pi^{-1} [ V \cos \pi z + \sin \pi x \cos \pi y - D \cos \pi x \sin \pi y \sin 2 \pi w ] \ .
\end{aligned}
\label{eq:Hamiltonians}
\end{equation}
Denote by $N_{xy}^h$, $N_{wx}^h$ and $N_{zy}^h$ a numerical integration scheme with step size $h$ for each of the Hamiltonian systems respectively.
Then, integration of \eqref{eq:HamiltonianDynamics} can be performed by a symmetric splitting method,
\begin{equation}
	N_{zy}^{h/2} N_{wx}^{h/2} N_{xy}^h N_{wx}^{h/2} N_{zy}^{h/2}.
\label{eq:splitting}
\end{equation}
In practice, we apply a Verlet method for each of the integrators. 
Since the Hamiltonians are not separable, the Verlet schemes are implicit.
However, the implicit equation in each step is scalar and is easily solved to machine precision with a few Newton-Raphson iterations.
In order  to make sure that the solution is consistent with the original 
system \eqref{eq:model}, we project the solution onto the $w=-z$ 
submanifold after each integration step, using the simple linear
projection  $z \leftarrow (z-w)/2$, $w \leftarrow (w-z)/2$.
Although the projection is not reversible, in practice, using 
this scheme with a step size of $h=10^{-3}$ or smaller 
is accurate to a high degree of  precision.

\subsection{Super diffusive properties}

\begin{figure}[t]
   \centerline{\includegraphics[width=0.8\textwidth]{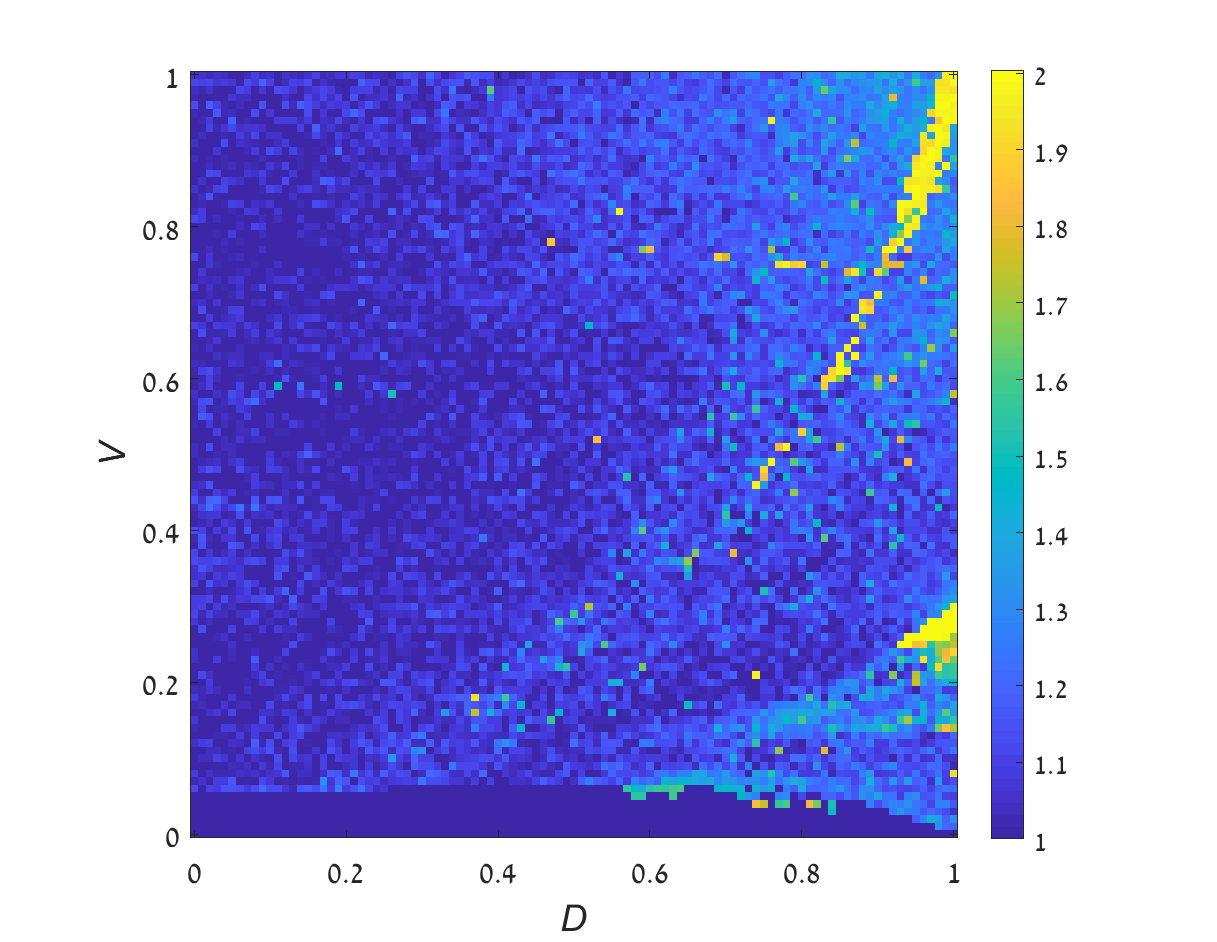}}
   \caption{Measured exponent of super-diffusivity as a function of $D$ and $V$}  
\label{fig:parameterScan}          
\end{figure}

The parameters $D$ and $V$  determine the geometric properties of trajectories.
One simple way to quantify how super-diffusive trajectories are is by calculating
the Mean Squared Displacement (MSD).
The physical motivation of the model is in describing transport or diffusion of particles 
in a flow.  Accordingly, we focus on diffusion  in the unwrapped $x$-$y$ plane, i.e.,
in $\R^2$ and not on the torus. 
Denoting by $r(t)=(x(t),y(t))$ the position in the $x$-$y$ plane at time $t$,
the MSD  is defined as
\begin{equation}
   {\rm MSD} (t) = < | r(t+s) - r(s) |^2 >   \ . 
\end{equation}
where $< \cdot >$ denotes a suitable average. This might be an average over all times 
$s$ for a specific orbit (in the chaotic region), or an average over all possible 
initial conditions $r(s)$ for some fixed $s$. Typically, 
\begin{equation}
   {\rm MSD} (t) \sim D t^\alpha   \ , 
\end{equation}
where  $0 \le \alpha \le 2$. $\alpha$ is called the exponent of diffusivity. 
Bounded trajectories have $\alpha=0$. 
In cases where the standard central limit theorem holds, we have regular diffusion with $\alpha=1$.
$\alpha=2$ implies ballistic motion with an effectively constant velocity.
Systems with $0<\alpha<1$ are called sub-diffusive and 
systems with  $1<\alpha<2$ are called super-diffusive.
Figure~\ref{fig:parameterScan} shows the 
numerically calculated  MSD exponent $\alpha$ as a function of $D$ and $V$. 
Values of $D$ and $V$ were taken in increments of $0.01$. 
For each value of $D$ and $V$
a single orbit of length $10^4$ was computed, and the MSD calculated, as an average
along the orbit, for a number of values of $t$, allowing the 
exponent of super-diffusivity
to be found by a linear fit of values of $\log{\rm MSD}(t)$ as a function of $\log t$. 

One conclusion that can be drawn from this plot is that 
in certain regions of parameter space we have $\alpha=2$, i.e. the dynamics
appears ballistic. This turns out to be a result of the existence of 
asymptotically stable periodic orbits, which will be explored further in the 
next subsection. The results are arguably 
consistent with the hypothesis that for $0<D,V<1$ we always have 
$\alpha>1$, i.e., trajectories are super-diffusive. However, the number of values of 
$\alpha$ observed above $1.4$ is small, and there is sensitive dependence on the 
values of $V$ and $D$, so it is not impossible that in fact $\alpha=1$ or $\alpha=2$
for all values of $V$ and $D$, c.f. the discussion in \cite{HC} and \cite{Klages}. 
It is evident that an orbit of length $10^4$ is not long enough to reliably compute 
the average defining the MSD. However, it is also clear that on time scales of 
this length, there is significant evidence of super-diffusive behavior. 

\subsection{Dissipative behavior}
\label{sec:GW}

\begin{figure}[t]
   \centerline{ \includegraphics[width=0.8\textwidth]{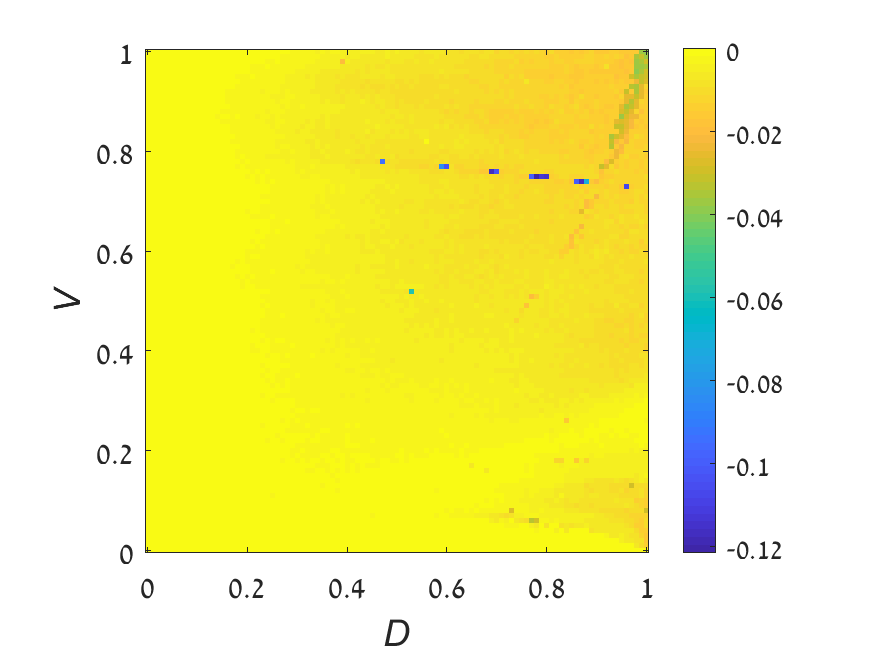} }
   \caption{Time average of the divergence of the flow along an orbit
      of length $10^4$ in the 
      chaotic region of flow, as a function of $D$ and $V$.}
   \label{fig:divcolor}
\end{figure}

As mentioned in the previous section,  in  Figure~\ref{fig:parameterScan} 
we see regions in which the exponent of super-diffusivity appears
to be $2$. Further numerical investigation of these regions shows that the reason 
for this is
that in these regions, there exist asymptotically stable periodic orbits, and, furthermore,
generic orbits outside the elliptic islands eventually 
tend to one of these attractors.  As we have noted previously, the divergence of  
the right hand side of    \eqref{eq:model} is 
$ - 2 \pi D \cos \pi x \cos \pi y \cos 2 \pi z$, which is not of fixed sign. If 
generic orbits tend to an attractor, we expect the time average of the divergence 
to be negative. In Figure  \ref{fig:divcolor}  we show the 
dependence on $D$ and $V$ of the time average of the
divergence along a single orbit of length $10^4$ in the chaotic region of the 
flow. As expected, we see regions in which the time average of the divergence is 
significantly negative. 

\begin{figure}[t]
   \centerline{ \includegraphics[width=0.9\textwidth]{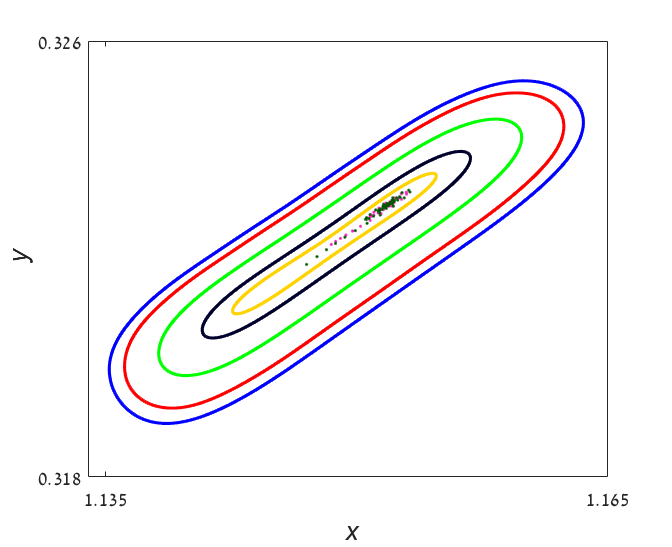} }
   \caption{Return map to $\theta=0.75$ for $V=0.6$, $D=0.865$. An asymptotically
   stable fixed point sits inside a family of invariant curves. The innermost curve 
   bounds the basin of attraction of the fixed point.}
   \label{fig:nestedislands}
\end{figure}

How do the attractors appear/disappear at the edges of the regions in 
parameter space in which they occur?  We studied the change in the attractor 
that exists for $V=0.6$, $D=0.84$  as $D$ is changed for fixed $V$. As $D$ is 
reduced (around $D=0.83$)  there is a period doubling bifurcation, 
which initiates  a period doubling cascade as $D$ is reduced further, leading to 
the disappearance of the periodic orbit. As $D$ is increased, at some stage 
(before $D=0.86$) the attractor loses the property of being a 
global attractor for generic orbits outside the elliptic islands. Instead, the
attractor is surrounded by an invariant surface, which is the 
boundary of its basin of attraction, and this is surrounded by other invariant surfaces. 
In figure \ref{fig:nestedislands} we show part of the return map to the 
surface of section $\theta=0.75$ for parameter values $V=0.6$ and $D=0.865$. 
A total of 7 orbits are shown. For 5 of these, the orbits fill a curve. For the 
remaining 2, the orbits tend to a fixed point at (approximately) $(1.15188,0.32300)$. 
As $D$ is increased even further, this fixed point changes from being asymptotically
stable to being neutrally stable, and an elliptic island appears (for the 
return map, the two eigenvalues go from being both real and smaller than $1$ in 
absolute value, to being a complex conjugate pair of absolute value $1$).
Approaching the bifurcation point, the area of the basin of attraction of the stable orbit seems to shrink to zero.

Section 5.2 of \cite{RobertsQuispel1992} describes the bifurcation of a neutrally stable 
symmetric fixed point in a reversible map of the plane to a hyperbolic fixed 
point and a symmetric, neutrally stable two-cycle. The observation is made 
that some of the ``outer'' invariant curves surrounding the original fixed point 
persist, and surround the hyperbolic fixed point and the two-cycle after 
the bifurcation. This is the situation observed in Figure 
\ref{fig:fractals}(d).  The hyperbolic fixed point in this figure 
(between the two sub-islands of the big central island) has 
one eigenvalue with absolute value less than $1$, and the other with absolute
value greater than $1$. The situation we see in figure \ref{fig:nestedislands}, 
however, is 
formed (as $D$ decreases, for fixed $V$) by the bifurcation of a neutrally stable 
{\em asymmetric} fixed point
to a hyperbolic stable fixed point, with two eigenvalues with absolute value 
less than $1$. Recall that in a reversible map, all symmetric fixed points 
are necessarily neutrally stable; but there can also exist asymmetric neutrally
stable fixed points. Also in the case of this bifurcation, some 
of the invariant curves surrounding the fixed point persist, as the type 
of the fixed point changes, giving rise to the picture in  figure~\ref{fig:nestedislands}. 
We are not aware that the occurrence of such a bifurcation has previously been 
reported. 

As a final note on this section of dissipative behavior, we mention that 
even in the case that all orbits outside outside the elliptic islands tend to an
attractor, resulting, ultimately, in ballistic motion, 
the typical time scale on which this happens is very long (of 
order $10^5$ in cases we have investigated). On shorter time scales the 
phenomenon of stickiness around the elliptic islands dominates. 

\subsection{Sticking times}
\label{sec:ST}

\begin{figure}[t]
\centerline{\includegraphics[width=0.5\textwidth]{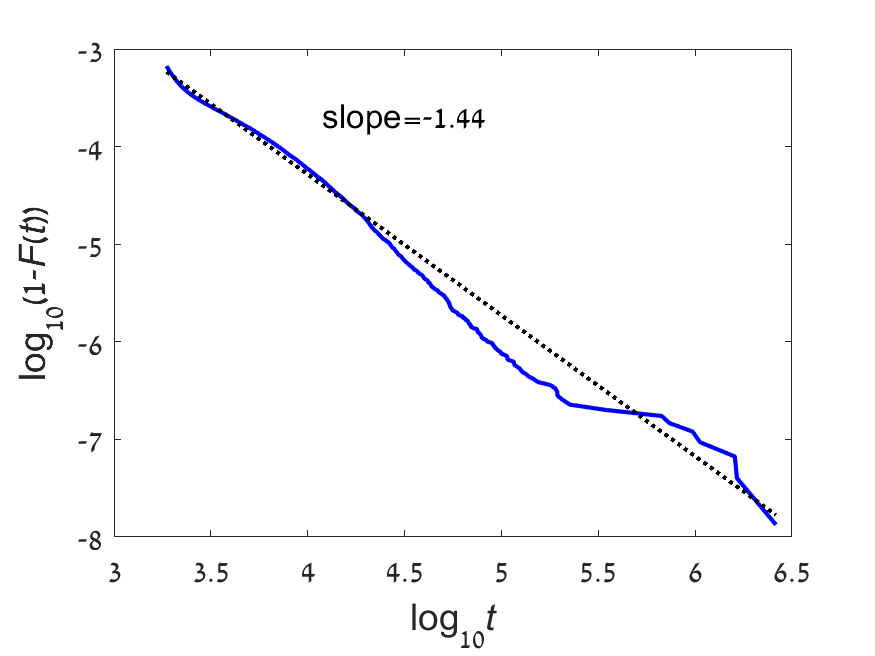}
~~\includegraphics[width=0.5\textwidth]{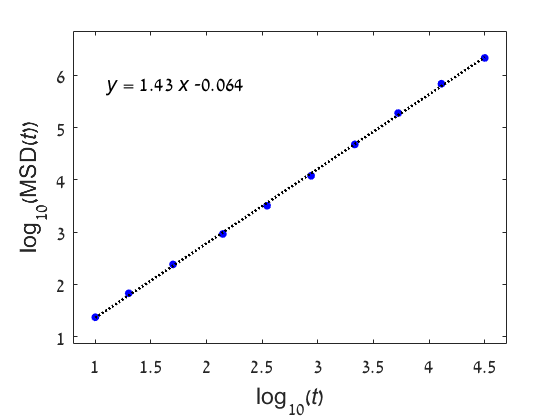}   }
\caption{
Left: The cumulative distribution function of sticking times around the regular islands 
has approximately a power-law tail with exponent $\gamma=1.44$.
Right: The MSD of the same orbit ($T=8\cdot 10^7$) shows super diffusion with exponent $\alpha\approx1.43$. The result is consistent with the standard L\'evy walk model, in which $\gamma=3-\alpha$ \cite{Zaburdeav2015}.}
\label{fig:noise}
\end{figure}

We report one final numerical study, on the distribution of sticking 
times. In the case that there are no attractors (and in the case that these
have bounded basins of attraction), we have argued that the dominant feature of the
dynamics outside the  elliptic islands is that orbits spend a long time 
``sticking'' to the islands. It is not clear precisely how to quantify this. 
We performed the following experiment: 
In the case $D=12/13$, $V=1/2$ we looked at a single very long orbit 
in the chaotic regime. We chose a rectangular region around the elliptic 
islands in the Poincar\'e section $z=-0.2$, i.e. 
the holes in figure~\ref{fig:sticky}. We recorded the lengths of the times
that the orbit stayed in this region, i.e. the ``sticking time'' 
and used these to build an 
empirical cumulative distribution function $F(t)$, that gives the probability 
that a given sticking time is less than $t$. We expect power law behavior for the
tail of this distribution, i.e. 
$$  1 -F(t) \sim Ct^{-\gamma} \ ,\qquad t\rightarrow\infty  $$
for some positive constants $C$ and $\gamma$. In Figure~\ref{fig:noise} 
we display a log-log plot of $1-F(t)$ against $t$ for large $t$ (corresponding 
to the $25,000$ longest sticking times). The 
result is consistent with power law behavior with $\gamma\approx1.44$. For a standard L\'evy walk, $\gamma=3-\alpha$ \cite{Zaburdeav2015}, which is consistent with Figure~\ref{fig:noise}b showing that $\alpha\approx1.4$.

\section{Summary}
\label{sec:summary}
We have analyzed the system \eqref{eq:model}
describing the position and orientation of 
an infinitesimal, self-propelled, prolate spheroid,  advected by an 
array of vortices. This is a non-Hamiltonian, non-volume preserving reversible 
system which, despite its superficial simplicity, displays a range of 
interesting dynamical properties. Most significantly, on intermediate and/or
long time scales the orbits have properties reminiscent of those of L\'evy
walks, and thus the system provides a framework for L\'evy walking in 
a deterministic setting. The mechanism for this is the existence of 
elliptic islands, (small) regions of phase space 
around neutrally stable periodic orbits
in which the motion is ballistic; 
these islands are ``sticky'' in the sense that other orbits spend long periods of 
time close to them, giving the long linear stretches characteristic of L\'evy walks. 
For certain values of the parameters the system can also
have attracting periodic orbits, which may even be global attractors, in the 
sense that all orbits outside the elliptic islands are attracted to them. 
This, however, only happens on very long time scales, and on intermediate
scales the L\'evy walk behavior is still observed. 

\bigskip

{\bf Acknowledgments}
We thank Or Alus, Rainer Klages and Ed Ott for discussions and suggestions. 
G.A. is thankful for partial support from The Israel Science Foundation 
(Grant No. 337/12), and DFG grant 1D-84024.

\end{document}